\documentclass{INTERSPEECH2023}

\usepackage{amsmath,graphicx}
\usepackage{cite}
\usepackage{algorithmic}
\usepackage{array}
\usepackage{stfloats}
\usepackage{xcolor}
\usepackage{epsfig,setspace}
\usepackage{float,algorithm}
\usepackage{bm}
\usepackage{multirow, hhline}
\usepackage{makecell}
\usepackage{amssymb}
\usepackage{pifont}
\usepackage{arydshln}
\usepackage{booktabs}
\newcommand{\cmark}{\ding{51}}
\newcommand{\xmark}{\ding{55}}
\newcommand{\tabincell}[2]{\begin{tabular}{@{}#1@{}}#2\end{tabular}}
\usepackage{subfig}
\usepackage{caption}
\usepackage{amsmath}

\usepackage[section]{placeins}

\usepackage{todonotes} 

\interspeechcameraready

\title{On-the-Fly Feature Based Rapid Speaker Adaptation\\ for Dysarthric and Elderly Speech Recognition}

\name{Mengzhe Geng$^{1,4,*}$, Xurong Xie$^2$, Rongfeng Su$^3$, Jianwei Yu$^4$, Zengrui Jin$^1$, Tianzi Wang$^1$, \\ Shujie Hu$^1$,Zi Ye$^1$, Helen Meng$^1$, Xunying Liu$^1$\vspace{-0.5em}\thanks{* Part of this work was done while the author was an intern at Tencent AI Lab.}}

\address{
  $^1$The Chinese University of Hong Kong, $^2$Institute of Software, Chinese Academy of Sciences \\
  $^3$Shenzhen Institute of Advanced Technology, Chinese Academy of Sciences,
  $^4$Tencent AI Lab\vspace{-1em}
}

\email{\{mzgeng,zrjin,twang,zye,hmmeng,xyliu\}@se.cuhk.edu.hk, xurong@iscas.ac.cn, rf.su@siat.ac.cn, tomasyu@tecent.com}

\begin{document}
\bstctlcite{IEEEexample:BSTcontrol}

\maketitle
 
\begin{abstract}

Accurate recognition of dysarthric and elderly speech remain challenging tasks to date. Speaker-level heterogeneity attributed to accent or gender, when aggregated with age and speech impairment, create large diversity among these speakers. Scarcity of speaker-level data limits the practical use of data-intensive model based speaker adaptation methods. To this end, this paper proposes two novel forms of data-efficient\textcolor{red}{, feature-based} on-the-fly speaker adaptation methods: variance-regularized spectral basis embedding (SVR) and spectral feature driven f-LHUC transforms. Experiments conducted on UASpeech dysarthric and DementiaBank Pitt elderly speech corpora suggest the proposed on-the-fly speaker adaptation approaches consistently outperform baseline iVector adapted hybrid DNN/TDNN and E2E Conformer systems by statistically significant WER reduction of 2.48\%-2.85\% absolute (7.92\%-8.06\% relative), and offline model based LHUC adaptation by 1.82\% absolute (5.63\% relative) respectively.

\end{abstract}


\section{Introduction}


Despite the breakthroughs in automatic speech recognition (ASR) technologies targeting normal speech, accurate recognition of dysarthric and elderly speech remains highly challenging tasks to date~\cite{christensen2012comparative,sehgal2015model,hu2019cuhk,xiong2020source,liu2020exploiting,geng2020investigation,hu2021bayesian,ye2021development,harvill2021synthesis,geng2021spectro,liu2021recent,jin2021adversarial,hu2022exploit,baskar2022speaker,wang2022conformer,geng2022spectro,yue2022acoustic,jin2022adversarial,hu2022exploring,geng2023use}. Speech impairments are commonly found among dysarthric speakers and the elderly experiencing natural aging and neurocognitive disorders~\cite{alzheimer20192019}. ASR technologies tailored to their needs can improve their quality of life.

Dysarthric and elderly speech presents a prominent challenge to current ASR technologies primarily targeting normal speech in many aspects. Heterogeneity commonly found in normal speech sourcing from accent or gender, when further combined with that over age and speech impairment severity, create large diversity among dysarthric and elderly speakers~\cite{kodrasi2020spectro,smith1987temporal}. Such diversity is further aggregated when spectral or temporal perturbation based data augmentation techniques ~\cite{vachhani2018data,xiong2019phonetic,geng2020investigation} are used. To this end, speaker adaptation techniques play a crucial role in the personalization of ASR systems for such users.


Speaker adaptation techniques for normal speech have been widely studied in three broad categories: 1) speaker-dependent (SD) auxiliary speaker embedding features~\cite{abdel2013fast,saon2013speaker,huang2015investigation}; 2) feature transformations \textcolor{red}{generating canonical} features at acoustic front-ends~\cite{digalakis1995speaker,seide2011feature}; 3) model based approaches \textcolor{red}{using} specially designed SD DNN parameters~\cite{neto1995speaker,zhang2015parameterised,swietojanski2016learning}.

In contrast, there are limited prior researches on dysarthric and elderly speaker adaptation. Earlier works were mainly conducted on HMM based ASR systems, including MLLR and MAP adaptation~\cite{baba2002elderly,mengistu2011adapting,christensen2012comparative,kim2013dysarthric} and their combination with speaker adaptive training (SAT)~\cite{sehgal2015model}, feature-space MLLR (f-MLLR) based SAT~\cite{bhat2016recognition} and regularized speaker adaptation via Kullback-Leibler (KL) divergence~\cite{kim2017regularized}. More recent researches applied model based adaptation to current DNN based ASR systems, including direct parameter fine-tuning based adaptation in both hybrid TDNN~\cite{xiong2020source,takashima2020two} and end-to-end RNN-T~\cite{Shor2019,green2021automatic} systems, LHUC~\cite{liu2021recent,ye2021development} and Bayesian speaker adaptation~\cite{deng2021bayesian}. Spectro-temporal basis embedding features (SBE) based offline adaptation via speaker-level averaging was studied in~\cite{geng2021spectro, geng2022spectro}.

One major issue associated with the prior researches above is the lack of suitable rapid, on-the-fly adaptation techniques targeting dysarthric and elderly speech. Such methods serve as dual-purpose solutions to handle not only the difficulty in collecting large quantities of data from such speakers with mobility issues that are essential for model based adaptation but also their latency issues due to multi-pass decoding and SD parameter estimation. The Bayesian model based adaptation using very limited speaker data~\cite{liu2021recent,xie2021bayesian,deng2021bayesian} only addressed the aforementioned data scarcity issue, but the latency problem remains unvisited. Similarly, the spectro-temporal deep embedding features~\cite{geng2021spectro} computed and averaged over all speaker-level data incurred latency and precluded the use of on-the-fly adaptation.

In order to address this issue, two novel forms of feature-based on-the-fly rapid speaker adaptation approaches are proposed in this paper. The first is based on speaker-level variance-regularized spectral basis embedding (SVR) features. An additional variance regularization term is included when training spectral basis embedding DNNs~\cite{geng2021spectro,geng2022spectro} to ensure speaker-level homogeneity of the resulting embedding features and thus allow them to be applied on the fly during test time adaptation. The second approach uses on-the-fly feature-based LHUC (f-LHUC) transforms conditioned on spectral features. Specially designed regression TDNN~\cite{xie2019fast} predicting speaker-level LHUC transforms are used to directly generate and apply such parameters during test time adaptation and thus resolve the latency due to multi-pass decoding. Experiments were conducted on the largest available and most widely used UASpeech~\cite{kim2008dysarthric} dysarthric and DementiaBank Pitt~\cite{becker1994natural} elderly speech datasets. Consistent performance improvements were obtained by our proposed on-the-fly speaker adaptation approaches using both hybrid DNN/TDNN and E2E Conformer~\cite{gulati2020conformer} systems.

The main contributions of the paper are summarized below: 


{\textbf 1)} This paper presents the first work of on-the-fly feature-based fast speaker adaptation targeting dysarthric and elderly speech.  In contrast, previous works using feature-based dysarthric and elderly speaker adaptation \textcolor{red}{uses all speaker-level data} and operate in batch mode~\cite{geng2021spectro, geng2022spectro} while those using model based adaptation not only require the usage of all speaker-level data, but also additional multiple decoding passes and explicit parameter estimation in test time~\cite{ geng2020investigation,ye2021development,deng2021bayesian,liu2021recent}. These prior works incur significant latency and are not the on-the-fly, rapid speaker adaptation approaches considered in this paper.


{\textbf 2)} Our proposed SVR features can instantaneously extract homogeneous dysarthric and elderly speaker characteristics on the fly. Experiments conducted on benchmark UASpeech dysarthric and DementiaBank Pitt elderly speech datasets suggest the proposed on-the-fly speaker adaptation approaches consistently outperform the baseline iVector adapted hybrid DNN/TDNN and E2E Conformer systems by statistically significant word error rate (WER) reduction of 2.48\%-2.85\% absolute (7.92\%-8.06\% relative), and offline model based LHUC adaptation by 1.82\% absolute (5.63\% relative) respectively.

\vspace{-0.5em}
\section{Variance-Regularized Spectral Features}
\label{sec-variance}
\begin{figure}[ht]
  \centering
  \setlength{\abovecaptionskip}{0.15cm}   
  \setlength{\belowcaptionskip}{-0.15cm}   
  \includegraphics[scale=0.62]{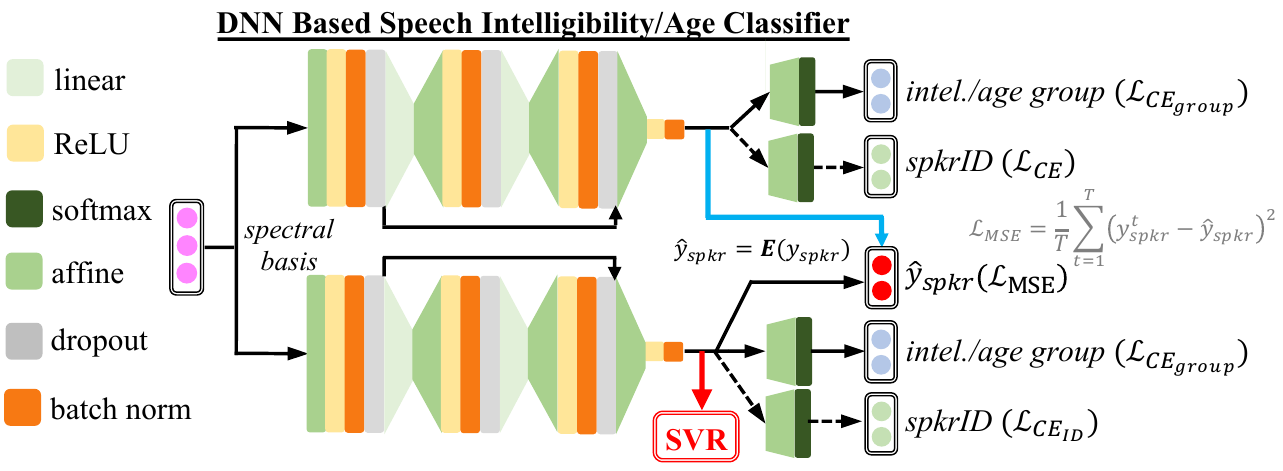}
  \caption{Our proposed DNN speech intelligibility or age classifier with a bottleneck layer to extract variance-regularized spectral basis embedding (SVR) features for speaker adaptation. Here ``intel.'' denotes speech intelligibility.}
  \label{fig:classifier}
\end{figure}

To model the latent diversity in dysarthric and elderly speech, singular value decomposition (SVD) is performed on Mel-filterbank log amplitude spectrum $\mathbf{S}$~\cite{van1993subspace}, given as:

\begin{equation}
\setlength{\abovedisplayskip}{-3pt}
\setlength{\belowdisplayskip}{-3pt}
    \mathbf{S = U{\Sigma}V^{\mathrm{T}}}
\end{equation}

where the top-$d$ principal spectral bases are retrieved from the column vectors of $\mathbf{U}$. Following~\cite{geng2021spectro,geng2022spectro}, further supervised learning is performed via constructing DNN speech intelligibility or age classifier (the upper part of Fig.\ref{fig:classifier}). The inputs are the selected spectral bases, and the targets are speech intelligibility groups $+$ speaker IDs for the UASpeech corpus and binary aged vs. non-aged annotations for the DementiaBank Pitt corpus.

To ensure the speaker-level homogeneity of the embedding features, a pair of such DNN classifiers (Fig.\ref{fig:classifier}) are constructed. The $25$-dim embedding features taken from the bottleneck layer of the upper classifier in Fig.~\ref{fig:classifier} are averaged by speaker and then serve as the regression targets of the lower DNN classifier (Fig.\ref{fig:classifier} blue bold line) for variance regularization. A multitask learning (MTL) cost function is used to train the lower classifier, using interpolation between 1) the cross-entropy (CE) error computed over speech intelligibility or age labels, optionally plus that computed over speaker IDs and 2) the mean squared error (MSE) computed between the lower DNN bottleneck features and the corresponding speaker-level averaged embedding features produced by the upper DNN, which is given as:

\begin{equation}
    \setlength{\abovedisplayskip}{-3pt}
    \setlength{\belowdisplayskip}{-3pt}
    \mathcal{L}_{MTL}={\omega_1}{\cdot}\mathcal{L}_{MSE}+{\omega_2}{\cdot}\mathcal{L}_{CE_{group}}+{\omega_3}{\cdot}\mathcal{L}_{CE_{ID}}\footnote{Empirically set for the UASpeech corpus as $\omega_1=\omega_2=\omega_3=\frac{1}{3}$, while for the DementiaBank Pitt corpus $\omega_1=\omega_2=\frac{1}{2},\omega_3=0$. }
\end{equation}

\begin{figure}[ht]
  \centering
  \vspace{-0.2cm} 
  \setlength{\abovecaptionskip}{0.15cm}   
  \setlength{\belowcaptionskip}{-0.35cm}   
  \includegraphics[scale=0.38]{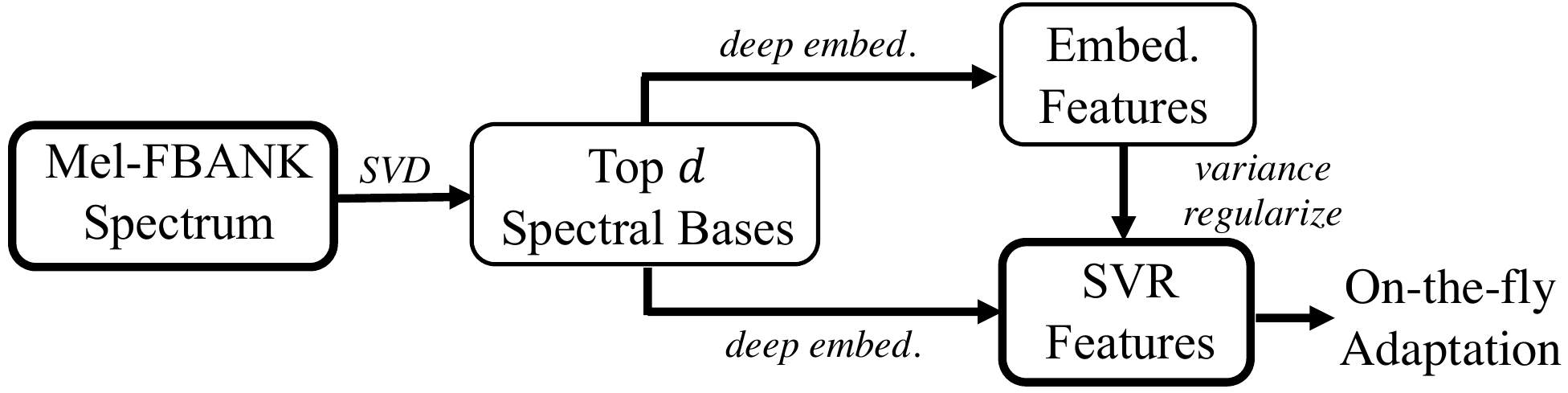}
  \caption{Procedure of generating the SVR features.}
  \label{fig:procedure}
\end{figure}

\begin{figure}[ht]
  \centering
  \vspace{-0.2cm} 
  \setlength{\abovecaptionskip}{0.2cm}   
  \setlength{\belowcaptionskip}{-0.35cm}   
  \includegraphics[scale=0.5]{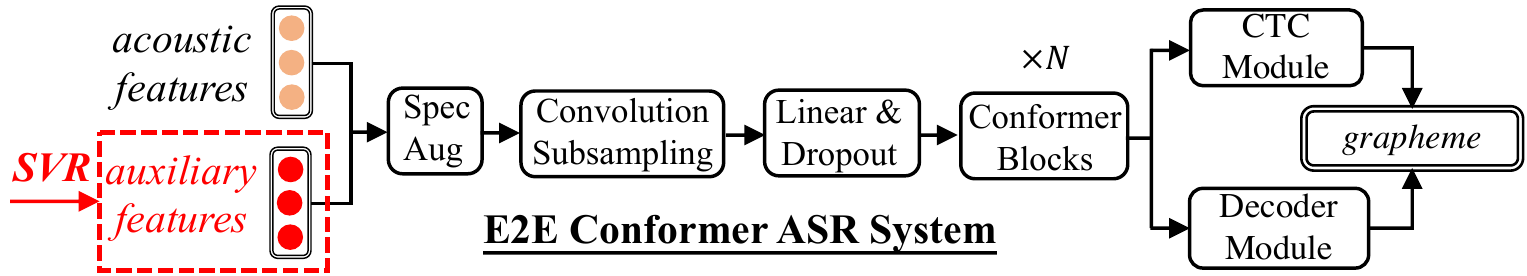}
  \caption{Incorporating SVR features into E2E Conformer.}
  \label{fig:E2E}
\end{figure}

The $25$-dim variance-regularized spectral basis embedding (SVR) features are then taken from the bottleneck layer of the lower classifier (Fig.\ref{fig:classifier} red bold line), and appended to the acoustic features at the front-end of hybrid DNN/TDNN (Fig.\ref{fig:system}) and E2E Conformer systems (Fig.\ref{fig:E2E}), to facilitate on-the-fly test time adaptation. The aforementioned procedure is summarized in Fig.~\ref{fig:procedure}. Directly using the top $d$ spectral bases or intermediate embedding features are unsuitable for speaker adaptation~\cite{geng2022spectro}.

\vspace{-0.5em}
\section{On-the-Fly F-LHUC Transforms}
\label{sec-lhuc}

\begin{figure}[ht]
  \centering
  \vspace{-0.2cm} 
  \setlength{\abovecaptionskip}{0.15cm}   
  \setlength{\belowcaptionskip}{-0.15cm}   
  \includegraphics[scale=0.68]{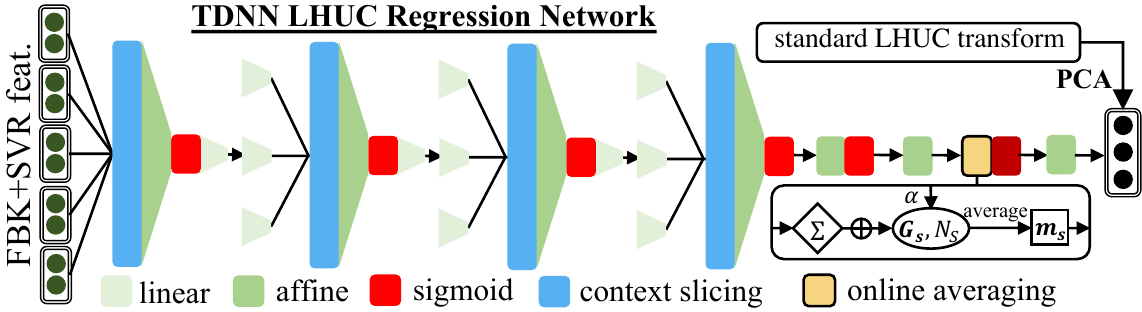}
  \caption{TDNN based LHUC regression network.}
  \label{fig:regression}
\end{figure}

In feature-based learning hidden unit contributions (f-LHUC) based adaptation approaches~\cite{xie2019fast}, LHUC transforms are predicted from the acoustic features on the fly. Supervised estimation of LHUC transforms on the training data is first conducted via standard speaker adaptive training (SAT). Principal component analysis (PCA) is further applied to produce compressed LHUC vectors encoding the most distinctive speaker-level features. These serve as the output targets for the TDNN based LHUC regression network (Fig.\ref{fig:regression}), with a specially designed online averaging layer~\cite{xie2019fast} given as:

\begin{equation}
    \setlength{\abovedisplayskip}{-3pt}
    \setlength{\belowdisplayskip}{-3pt}
    \bm{m_{s}^{i}}=\frac{\bm{\sum_{t=1}^{T_i}\bm{h_{t}^{i}}+{\alpha}{\cdot}\bm{G_{s}^{i-1}}}}{T_{i}+{\alpha}{\cdot}N_{s}^{i-1}}
\end{equation}


\begin{figure}[ht]
  \centering
  \vspace{-0.2cm} 
  \setlength{\abovecaptionskip}{0.15cm}   
  \setlength{\belowcaptionskip}{-0.35cm}   
  \includegraphics[scale=0.54]{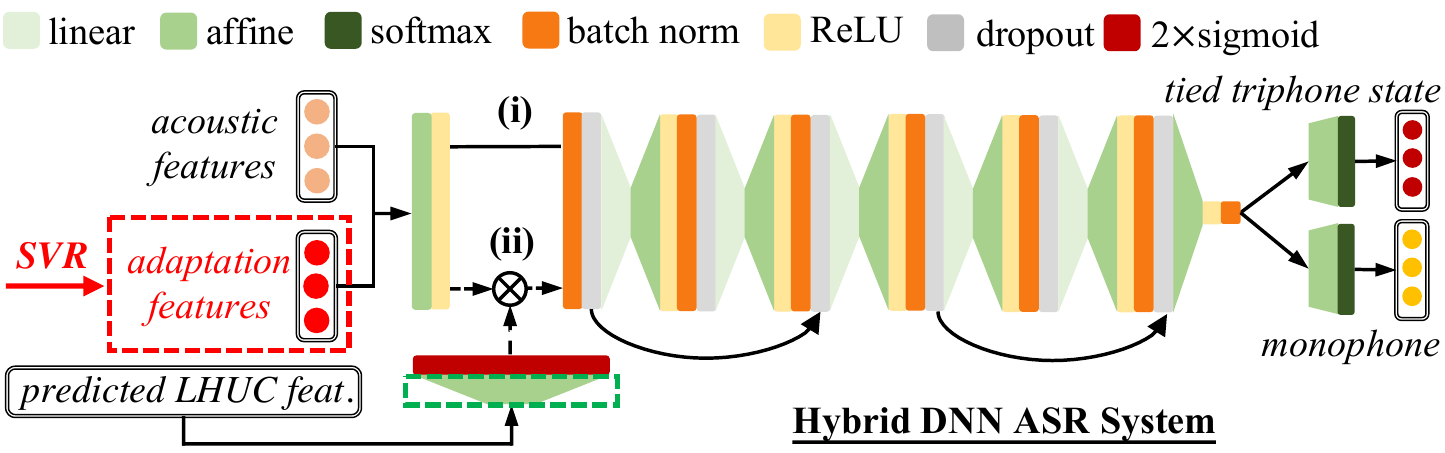}
  \caption{Incorporation of variance-regularized spectral basis embedding (SVR) features at front-end of hybrid DNN ASR systems~\cite{liu2021recent}. Selecting path (i) leads to systems with auxiliary feature based adaptation only, while selecting (ii) leads to systems with additional feature-based LHUC (f-LHUC) adaptation. }
  \label{fig:system}
\end{figure}

where \bm{$G_s$}, $N_s$ and \bm{$m_s$} denote the accumulated hidden vector, frame count, and averaged hidden vector till the $i^{th}$ segment of speaker $s$. The $i^{th}$ \textcolor{red}{utterance} contains $T_{i}$ frames, and the hidden vector of the $t^{th}$ frame is $\bm{h_{t}^{i}}$. $\alpha \in [0,1]$ is the history interpolation weight. Different from~\cite{xie2019fast} where only Mel-filterbank (FBK) features are considered, FBK $+$ variance-regularized spectral basis embedding (SVR) features are used to train the LHUC regression network (Fig~\ref{fig:regression}). An additional affine transformation is further trained to map the predicted low-dimensional LHUC features for training speakers to their corresponding LHUC transforms. During test time on-the-fly adaptation, the regression network (Fig.~\ref{fig:regression}) and the affine transformation (Fig.\ref{fig:system} circled in green) are applied in turn to generate speaker-level LHUC transforms using FBK $+$ SVR features.

\vspace{-0.5em}
\section{Experiments and Results}
\label{sec-exp}

\subsection{Experiments on the UASpeech Dataset}
\label{sec-exp-uaspeech}

\textbf{Task Description:} The UASpeech dataset is the largest publicly available and widely used dysarthric speech dataset~\cite{kim2008dysarthric}, which is an isolated word recognition task containing $103$h speech from $16$ dysarthric and $13$ control speakers. It is split into three blocks B1, B2 and B3, each with the same $155$ common words and a different set of $100$ uncommon words. The training set includes B1 and B3 data of all $29$ speakers ($69.1$h), while the test set includes B2 data of $16$ dysarthric speakers ($22.6$h, excluding speech from control speakers). Silence stripping using an HTK~\cite{young2002htk} trained GMM-HMM system~\cite{liu2021recent} produces a $30.6$h training set ($99195$ utt.) and a $9$h test set ($26520$ utt.). Data augmentation~\cite{geng2020investigation} produces a $130.1$h augmented training set ($399110$ utt.). The average utterance length is $1.2$ seconds. As E2E systems are sensitive to the training data coverage, B2 data of the $13$ control speakers and their augmented version via speed perturbation are also used for Conformer system training. This creates a $190$h training set ($538292$ utt.).


\noindent\textbf{Experiment Setup:} The $7$-layer hybrid DNN and E2E graphemic Conformer systems were implemented using Kaldi~\cite{povey2011kaldi} following ~\cite{liu2021recent} and ESPnet~\cite{watanabe2018espnet}\footnote{$12$ encoder layers + $12$ decoder layers, feed-forward dim = $2048$, $4$ attention heads of $256$ dimensions, interpolated CTC+AED cost.}. The inputs to both systems were $80$-dim filter-bank (FBK) + $\Delta$ features plus $25$-dim variance-regularized spectral basis embedding (SVR) features or $100$-dim iVectors\footnote{Kaldi: egs/wsj/s5/local/nnet3/run\_ivector\_common.sh. Changing the dimensionality of iVectors produces margin effect~\cite{geng2022spectro}.} (Fig.~\ref{fig:system}-~\ref{fig:E2E}). Top $2$ principal spectral bases~\cite{geng2021spectro,geng2022spectro} were used to train the DNN speech intelligibility classifier (Fig.\ref{fig:classifier}). The history interpolation weight $\alpha$ of the LHUC regression network was set to $0.9$ with the four context slicing indices as $\{-2,-1,0,1,2\}$,$\{-2,0,2\}$,$\{-3,0,3\}$ and $\{-4,0,4\}$. A uniform language model (LM) was used in decoding~\cite{christensen2012comparative}. As an ablation study, we used iVectors as the inputs to the DNN classifier (Fig.\ref{fig:classifier}) for variance regularization and generate variance-regularized iVectors (iVRs).

\begin{table}[ht]
    \caption{Performance of the proposed variance-regularized spectral basis embedding (SVR) feature adaptation, iVector adaptation and LHUC adaptation on the \textbf{UASpeech} test set of $16$ dysarthric speakers. ``SBE'' denotes spectral basis embedding features. ``VL/L/M/H" refer to intelligibility very low, low, mid and high. ``On Fly'' indicates using on-the-fly adaptation. $^\dag$ denotes a statistically significant improvement ($\alpha=0.05$) is obtained over iVector adapted systems (Sys.2,12,23).}
    \label{tab:recog-UASpeech}
    \Large
    \centering
    \vspace{-0.5em}
    \setlength{\abovecaptionskip}{0.05cm}
    \renewcommand\arraystretch{1.0}
    \renewcommand\tabcolsep{2.0pt}
    \scalebox{0.39}{\begin{tabular}{c|c|c|c|c|c|c|c|cccc|c}
    \hline\hline  
        \multirow{2}{*}{Sys.} & 
        \multirow{2}{*}{\tabincell{c}{Model\\(\#Para.)}} & 
        \multirow{2}{*}{\tabincell{c}{Data\\Aug.}} &
        \multirow{2}{*}{\#Hrs} &
        \multirow{2}{*}{\tabincell{c}{Adapt.\\Feat.}} & 
        \multirow{2}{*}{\tabincell{c}{LHUC\\SAT}} &
        \multirow{2}{*}{f-LHUC} &
        \multirow{2}{*}{\tabincell{c}{On\\Fly}} &
        \multicolumn{5}{c}{WER\%} \\
    \cline{9-13}
       & & & & & & & & VL & L & M & H & All \\
    \hline\hline
        1 & \multirow{11}{*}{\tabincell{c}{Hybrid\\DNN\\(6M)}} & \multirow{11}{*}{\xmark} & \multirow{11}{*}{30.6} & \xmark & \multirow{5}{*}{\xmark} & \multirow{5}{*}{\xmark} & - & 69.82 & 32.61 & 24.53 & 10.40 & 31.45 \\
        2 & & & & iVector & & & \cmark & 69.46 & 33.78 & 22.58 & 10.45 & 31.33 \\
        3 & & & & SBE~\cite{geng2021spectro} & & & \xmark & 64.43 & 29.71 & 19.84 & 8.57 & 28.05 \\
        4 & & & & iVR & & & \cmark & 68.66 & 33.72 & 22.84 & 9.83 & 30.99 \\
        5 & & & & SVR & & & \cmark & 65.04$^\dag$ & 30.90$^\dag$ & 20.70$^\dag$ & 10.15$^\dag$ & \textbf{28.85$^\dag$} \\
    \cline{1-1}\cline{5-13}
        6 & & & & \xmark & \multirow{2}{*}{\cmark} & \multirow{2}{*}{\xmark} & \xmark & 64.39 & 29.88 & 20.27 & 8.95 & 28.29 \\
        7 & & & & SBE~\cite{geng2021spectro} & & & \xmark & 63.40 & 28.90 & 18.64 & 8.13 & 27.24 \\
    \cline{1-1}\cline{5-13}
        8 & & & & \xmark & \multirow{3}{*}{\xmark} & (FBK) & \cmark & 66.47$^\dag$ & 29.55$^\dag$ & 21.00$^\dag$ & 8.99$^\dag$ & 28.80$^\dag$ \\
        9 & & & & iVector & & (+iVector) & \cmark & 64.86 & 36.44 & 21.17 & 9.03 & 30.29 \\
        10 & & & & SVR & & (+SVR) & \cmark & 65.75$^\dag$ & 29.80$^\dag$ & 19.07$^\dag$ & 8.99$^\dag$ & \textbf{28.31$^\dag$} \\
    \cline{1-1}\cline{5-13}
        5+10 & & & & \multicolumn{3}{c|}{-} & \cmark & \textbf{64.36$^\dag$} & \textbf{29.68$^\dag$} & \textbf{18.96$^\dag$} & \textbf{8.89$^\dag$} & \textbf{27.96$^\dag$}\\
    \hline\hline
        11 & \multirow{11}{*}{\tabincell{c}{Hybrid\\DNN\\(6M)}} & \multirow{11}{*}{\cmark} & \multirow{11}{*}{130.1} & \xmark & \multirow{5}{*}{\xmark} & \multirow{5}{*}{\xmark} & - & 66.45 & 28.95 & 20.37 & 9.62 & 28.73 \\
        12 & & & & iVector & & & \cmark & 65.73 & 30.10 & 20.21 & 9.03 & 28.65 \\
        13 & & & & SBE~\cite{geng2021spectro} & & & \xmark & 61.55 & 27.52 & 17.31 & 8.22 & 26.26 \\
        14 & & & & iVR & & & \cmark & 66.02 & 29.52 & 19.56 & 9.32 & 28.53 \\
        15 & & & & SVR & & & \cmark & 62.54$^\dag$ & 30.22 & 18.54$^\dag$ & 8.59$^\dag$ & \textbf{27.54$^\dag$} \\
    \cline{1-1}\cline{5-13}
        16 & & & & \xmark & \multirow{2}{*}{\cmark} & \multirow{2}{*}{\xmark} & \xmark & 62.50 & 27.26 & 18.41 & 8.04 & 26.55 \\
        17 & & & & SBE~\cite{geng2021spectro} & & & \xmark & 59.83 & 27.16 & 16.80 & 7.91 & 25.60 \\
    \cline{1-1}\cline{5-13}
        18 & & & & \xmark & \multirow{3}{*}{\xmark} & (FBK) & \cmark & 65.06 & 27.94$^\dag$ & 18.76$^\dag$ & 8.39$^\dag$ & 27.45$^\dag$ \\
        19 & & & & iVector & & (+iVector) & \cmark & 63.63 & 32.56 & 18.52 & 8.31 & 28.28 \\
        20 & & & & SVR & & (+SVR) & \cmark & 61.56$^\dag$ & 28.81 & 18.39$^\dag$ & 8.50$^\dag$ & \textbf{26.90$^\dag$} \\
    \cline{1-1}\cline{5-13}
        15+20 & & & & \multicolumn{3}{c|}{-} & \cmark &\textbf{60.80$^\dag$} &\textbf{28.19$^\dag$} &\textbf{17.72$^\dag$}  & \textbf{8.23$^\dag$} & \textbf{26.36$^\dag$} \\
    \hline\hline
        21 & \multirow{6}{*}{\tabincell{c}{Conformer\\(52M)}} & \multirow{6}{*}{\cmark} & 130.1 & \xmark & \multirow{6}{*}{\xmark} & \multirow{6}{*}{\xmark} & - & 73.88 & 53.12 & 49.92 & 42.03 & 53.17 \\
    \cline{4-5}\cline{8-13}
        22 & & & \multirow{5}{*}{190} & \xmark & & & - & 65.70 & 40.63 & 33.39 & 9.53 & 34.07 \\
        23 & & & & iVector & & & \cmark & 69.05 &  42.45 & 33.60 & 9.74 & 35.37 \\
        24 & & & & SBE~\cite{geng2021spectro} & & & \xmark & 65.18 & 34.90 & 24.21 & 5.00 & 29.19 \\
        25 & & & & iVR & & & \cmark & 68.94 & 42.00 & 32.19 & 8.52 & 34.55 \\
        26 & & & & SVR & & & \cmark & 67.52$^\dag$ & \textbf{38.85$^\dag$} & \textbf{28.60$^\dag$} & \textbf{7.88$^\dag$} & \textbf{32.52$^\dag$} \\
    \hline\hline
    \end{tabular}}
    \vspace{-6mm}
\end{table}


\noindent \textbf{Result Analysis:} Table~\ref{tab:recog-UASpeech} shows the performance comparison\footnote{A matched pairs sentence-segment word error (MAPSSWE) based statistical signiﬁcance test~\cite{pallet1990tools} was done at signiﬁcance level $\alpha=0.05$.} between the proposed variance-regularized spectral basis embedding (SVR) feature adaptation, spectral feature driven f-LHUC adaptation, iVector adaptation and offline LHUC adaptation on the UASpeech corpus. Several trends can be observed: \textbf{1)} On-the-fly SVR adaptation (Sys.5,15,26) consistently and statistically significantly outperform iVector adaptation (Sys.2,12,23) with various amounts of training data by up to \textbf{2.48\% absolute (7.92\% relative)} overall WER reduction for hybrid DNN (Sys.5 vs. Sys.2), and \textbf{2.85\% absolute (8.06\% relative)} reduction (Sys.26 vs. Sys.23) for Conformer systems, respectively. \textbf{2)} The improvements from offline LHUC adaptation (Sys.6,16) over the SI systems (Sys.1,11) are largely retained (by 82\%) and comparable to those obtained using on-the-fly SVR adaptation (Sys.5,15). \textbf{3)} Compared with improvements over SI systems (Sys.1,11) obtained by offline SBE adaptation~\cite{geng2021spectro} that requires expensive speaker-level averaging (Sys.3,13), on-the-fly SVR adaptation (Sys.5,15) produces comparable performance. \textbf{4)} The spectral feature (FBK+SVR) driven f-LHUC adapted systems (Sys.10,20) outperform both the FBK driven f-LHUC adapted systems (Sys.8,18) and the SVR adaptation alone (Sys.5,15). \textbf{5)} Frame-level log-likelihood score combination between the on-the-fly SVR adaptation and FBK+SVR driven f-LHUC adaptation leads to further improvements (Sys.5+10, Sys.15+20). \textbf{6)} The SVR on-the-fly adapted systems (Sys.5,15,26) consistently outperform comparable variance-regularized iVector (iVR) adaptation (Sys.4,14,25). \textbf{7)} Our proposed FBK+SVR driven f-LHUC adapted systems (Sys.10,20) consistently outperform the comparable FBK+iVector driven f-LHUC adaptation (Sys.9,19). \textbf{8)} A comparison between published systems on UASpeech and ours is shown in Table \ref{tab:compare}. Our best-performing system (Table~\ref{tab:recog-UASpeech}, Sys.15+20) produced the lowest WERs among all systems using online speaker adaptation in Table~\ref{tab:compare}.


\subsection{Experiments on the DementiaBank Pitt Dataset}
\label{sec-exp-dbank}
\textbf{Task Description:} The DementiaBank Pitt~\cite{becker1994natural} dataset contains $33$h speech recorded over interviews between $292$ elderly participants and clinical investigators. After split of the data and silence stripping~\cite{ye2021development}, the training set contains $15.7$h speech from $244$ elderly and $444$ investigators ($29682$ utt.) while the development and evaluation sets contain $2.5$h ($5103$ utt.) and $0.6$h ($928$ utt.) speech from $43$ elderly and $76$ investigators\footnote{The evaluation set is based on exactly the same $48$ speakers' Cookie task recordings following~\cite{luz2020alzheimer} while the development set contains the recordings of these speakers in other tasks if available.}. Data augmentation~\cite{ye2021development} produced an $58.9$h augmented training set ($112830$ utt.). The average utterance length is $1.9$ seconds.

\noindent\textbf{Experiment Setup:} The inputs to the hybrid TDNN systems\footnote{$14$ context slicing layers with a $3$-frame context~\cite{povey2011kaldi}.} and E2E graphemic Conformer systems\footnote{$12$ encoder layers + $12$ decoder layers, feed-forward dim = $2048$,  $4$ attention heads of $256$ dimensions, interpolated CTC+AED cost.} were $40$-dim FBK + $25$-dim SVR features or $100$-dim iVectors. Top $3$ spectral bases~\cite{geng2022spectro} served as the inputs to the DNN age classifier (Fig.\ref{fig:classifier}). A word $4$-gram LM~\cite{ye2021development} and a $3.8$k recognition vocabulary covering all words in DementiaBank Pitt corpus was used.

\vspace{-0.5em}
\begin{table}[ht]
    \caption{Performance of the proposed variance-regularized spectral basis embedding (SVR) adaptation, iVector adaptation and LHUC adaptation on the augmented \textbf{DementiaBank Pitt} corpus. ``INV'' and ``PAR'' denote investigator and elderly. ${\dag}$ denotes a statistically significant improvement ($\alpha=0.05$) over both the iVector adaptation (Sys.2) and offline LHUC adaptation (Sys.6), while ${\ddag}$ denotes a statistically significant improvement ($\alpha=0.05$) over the iVector adaptation (Sys.12).}
    \label{tab:recog-DBANK}
    \Large
    \centering
    \vspace{-0.5em}
    \setlength{\abovecaptionskip}{0.05cm}
    \renewcommand\arraystretch{1.0}
    \renewcommand\tabcolsep{2.0pt}
    \scalebox{0.42}{\begin{tabular}{c|c|c|c|c|c|c|cc|cc|c}
    \hline\hline  
        \multirow{3}{*}{Sys.} & 
        \multirow{3}{*}{\tabincell{c}{Model\\(\#Para.)}} & 
        \multirow{3}{*}{\#Hrs} &
        \multirow{3}{*}{\tabincell{c}{Adapt.\\Feat.}} & 
        \multirow{3}{*}{\tabincell{c}{LHUC\\SAT}} &
        \multirow{3}{*}{f-LHUC} &
        \multirow{3}{*}{\tabincell{c}{On\\Fly}} &
        \multicolumn{5}{c}{WER\%} \\  
    \cline{8-12}
        & & & & & & & \multicolumn{2}{c|}{Dev} & \multicolumn{2}{c|}{Eval} & \multirow{2}{*}{All} \\
    \cline{8-11}
        & & & & & & & INV & PAR & INV & PAR &  \\
    \hline\hline
        1 & \multirow{11}{*}{\tabincell{c}{Hybrid\\TDNN\\(18M)}} & \multirow{11}{*}{58.9} & \xmark & \multirow{5}{*}{\xmark} & \multirow{5}{*}{\xmark} & - & 19.91 & 47.93 & 19.76 & 36.66 & 33.80 \\
        2 & & & iVector & & & \cmark & 19.97 & 46.76 & 18.20 & 37.01 & 33.37  \\
        3 & & & SBE~\cite{geng2022spectro} & & & \xmark & 18.61 & 43.84 & 17.98 & 33.82 & 31.12  \\
        4 & & & iVR & & & \cmark & 19.19 & 47.64 & 18.65 & 35.80 & 33.26 \\
        5 & & & SVR & & & \cmark & 18.72$^{\dag}$ & 44.67$^{\dag}$ & 18.65 & 34.03$^{\dag}$ & \textbf{31.55$^{\dag}$}  \\
    \cline{1-1}\cline{4-12}
        6 & & & \xmark & \multirow{2}{*}{\cmark} & \multirow{2}{*}{\xmark} & \xmark & 19.26 & 45.49 & 18.42 & 35.44 & 32.33 \\
        7 & & & SBE~\cite{geng2022spectro} & & & \xmark & 17.41 & 40.94 & 17.98 & 31.89 & 29.16 \\
    \cline{1-1}\cline{4-12}
        8 & & & \xmark & \multirow{3}{*}{\xmark} & (FBK) & \cmark & 19.61 & 45.40 & 18.87 & 34.77 & 32.33 \\
        9 & & & iVector & & (+iVector) & \cmark & 18.75 & 47.07 & 17.98 & 36.11 & 32.85 \\
        10 & & & SVR & & (+SVR) & \cmark & 17.87$^{\dag}$ & 43.83$^{\dag}$ & 16.87$^{\dag}$ & 34.56$^{\dag}$ & \textbf{30.91$^{\dag}$} \\
    \cline{1-1}\cline{4-12}
        5+10 & & & \multicolumn{3}{c|}{-} & \cmark & 17.66$^{\dag}$ & \textbf{43.48$^{\dag}$} & 16.09$^{\dag}$ & \textbf{33.68$^{\dag}$} & \textbf{30.51$^{\dag}$} \\
    \hline\hline 
        11 & \multirow{5}{*}{\tabincell{c}{Conformer\\(52M)}} & \multirow{5}{*}{58.9} & \xmark & \multirow{5}{*}{\xmark} & \multirow{5}{*}{\xmark} & - & 20.97 & 48.71 & 19.42 & 36.93 & 34.57 \\
        12 & & & iVector & & & \cmark & 21.48 & 48.32 & 17.42 & 37.79 & 34.71  \\
        13 & & & SBE~\cite{geng2022spectro} & & & \xmark & 20.44 & 47.70 & 17.31 & 36.11 & 33.76  \\
        14 & & & iVR & & & \cmark & 22.09 & 49.56 & 19.64  & 38.58 & 35.65 \\
        15 & & & SVR & & & \cmark & 20.83 & \textbf{47.39$^\ddag$} & 17.64 & \textbf{36.34$^\ddag$} & \textbf{33.84$^\ddag$} \\
    \hline\hline     
    \end{tabular}}
\end{table}


\noindent \textbf{Result Analysis:} Table~\ref{tab:recog-DBANK} shows the performance of the proposed on-the-fly SVR adaptation, f-LHUC adaptation, iVector adaptation and LHUC adaptation on DementiaBank Pitt. Trends similar to those on UASpeech in Table~\ref{tab:recog-UASpeech} are observed: \textbf{1)} On-the-fly SVR adaptation (Sys.5,15) statistically significantly outperform iVector adaptation (Sys.2,12) on both TDNN and Conformer systems by up to \textbf{1.82\% absolute (5.45\% relative)} WER reduction (Sys.5 vs. Sys.2). \textbf{2)} On-the-fly SVR adaptation outperforms offline LHUC adaptation by 0.78\% absolute (2.41\% relative) overall WER reduction (Sys.5 vs. Sys.6). \textbf{3)} On-the-fly SVR adaptation (Sys.5) largely retains (by 84\%) the improvements obtained by offline SBE adaptation~\cite{geng2022spectro} (Sys.3) over the SI system (Sys.1). \textbf{4)} The proposed FBK+SVR driven f-LHUC adapted system (Sys.10) outperforms offline LHUC adaptation (Sys.6) while also outperforming FBK driven f-LHUC adaptation (Sys.8) and SVR adaptation alone (Sys.5). \textbf{5)} Frame-level score combination between SVR and FBK+SVR driven f-LHUC on-the-fly adaptation leads to \textbf{1.82\% absolute (5.63\% relative)} WER reduction over the offline LHUC adaptation (Sys.5+10 vs. Sys.6). \textbf{6)} Our proposed on-the-fly SVR (Sys.5,15) and FBK+SVR driven f-LHUC adaptation (Sys.10) consistently outperform the comparable iVR (Sys.4,14) and FBK+iVector driven f-LHUC adaptation (Sys.9).


\subsection{Further Ablation Studies}

\begin{figure}[ht]
  \centering
  \vspace{-0.2cm} 
  \setlength{\abovecaptionskip}{0.2cm}   
  \setlength{\belowcaptionskip}{-0.55cm}   
  \includegraphics[scale=0.56]{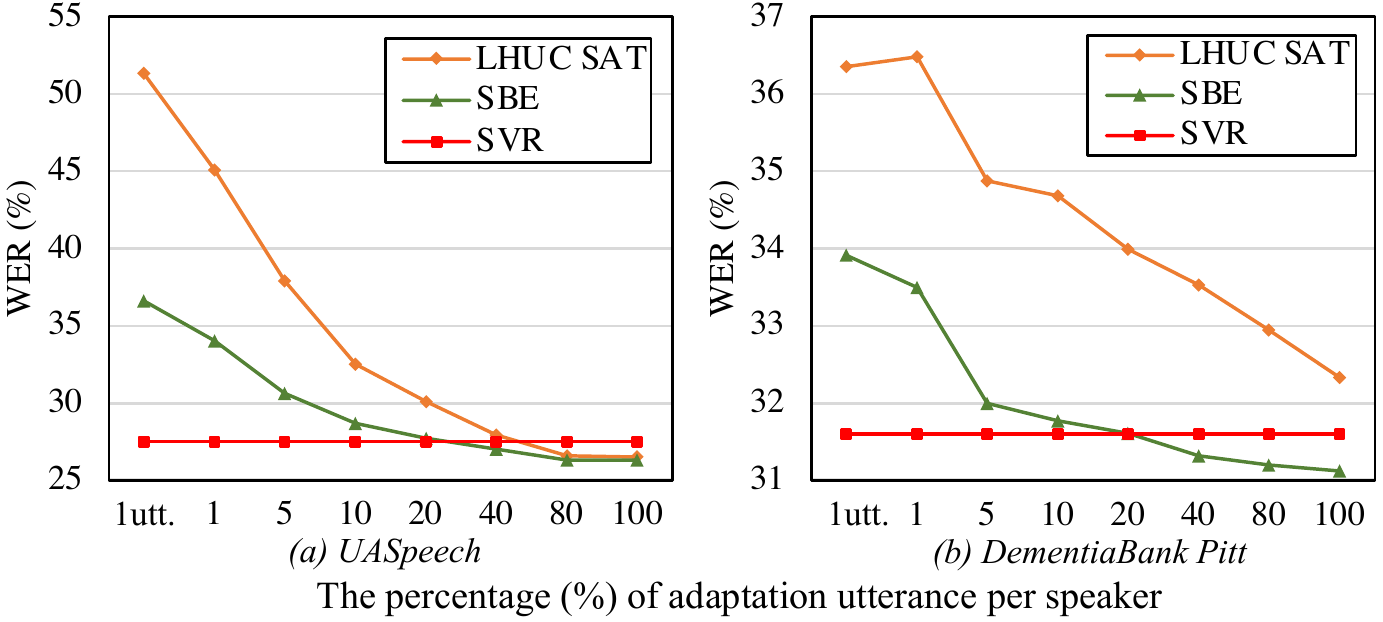}
  \caption{Performance (WER\%) of offline LHUC~\cite{liu2021recent}, offline SBE adaptation~\cite{geng2021spectro,geng2022spectro} and on-the-fly SVR adaptation of Sec.\ref{sec-variance} on varying percentage of test set speaker-level adaptation data on \textbf{UASpeech} and \textbf{DementiaBank Pitt}.}
  \label{fig:limit}
\end{figure}

As expected, the ablation study in Fig.~\ref{fig:limit} confirms on-the-fly SVR adaptation is more robust to varying amounts of speaker-level data used in adaptation than offline LHUC and SBE adaptation, and consistently outperforms both when less than 40\% of speaker-level data is used. A further ablation study in Table~\ref{tab:ablation} suggests the performance of on-the-fly SVR adaptation is largely insensitive to the length of analysis sliding windows (from 1 utt. down to 10ms). Homogeneous dysarthric and elderly speaker characteristics can be instantaneously extracted in the SVR features on the fly. The real-time (R.T.) factor indicates the total delay of waiting for data and model processing.



\vspace{-0.5em}
\begin{table}[ht]
    \caption{Ablation study on the augmented \textbf{UASpeech} (UA.) and \textbf{DementiaBank Pitt} (DBK.) corpora with various sizes of sliding window (Slid. Wind.) for on-the-fly SVR feature extraction. ``R.T.'' is short for real time. $^\dag$ denotes a statistically significant improvement ($\alpha=0.05$) is obtained over the comparable on-the-fly iVector adapted systems.}
    \label{tab:ablation}
    \Large
    \centering
    \vspace{-0.5em}
    \setlength{\abovecaptionskip}{0.05cm}
    \renewcommand\arraystretch{1.0}
    \renewcommand\tabcolsep{2.0pt}
    \scalebox{0.45}{\begin{tabular}{c|c|c|c|c|c|c|cccc|c}
    \hline\hline  
        \multirow{2}{*}{\tabincell{c}{UA.\\Sys.}} & 
        \multirow{2}{*}{\tabincell{c}{Model\\(\#Para.)}} & 
        \multirow{2}{*}{\#Hrs} &
        \multirow{2}{*}{\tabincell{c}{Adapt.\\Feat.}} & 
        \multirow{2}{*}{\tabincell{c}{Slid.\\Wind.}} &
        \multirow{2}{*}{\tabincell{c}{On\\Fly}} &
        \multirow{2}{*}{\tabincell{c}{R.T.\\Factor}} &
        \multicolumn{5}{c}{WER\%} \\
    \cline{8-12}
         & & & & & & & VL & L & M & H & All \\
    \hline\hline
        1 & \multirow{7}{*}{\tabincell{c}{Hybrid\\DNN\\(6M)}} & \multirow{7}{*}{130.1} & iVector & 100ms & \multirow{7}{*}{\cmark} & 0.10 & 65.73 & 30.10 & 20.21 & 9.03 & 28.65 \\
    \cline{1-1} \cline{4-5} \cline{7-12}
        2 & & & \multirow{7}{*}{SVR} & utt. & & 1.02 & 62.54$^\dag$ & 30.22 & 18.54$^\dag$ & 8.59$^\dag$ & 27.54$^\dag$ \\
        3 & & & & 300ms & & 0.27 & 63.74$^\dag$ & 29.01$^\dag$ & 19.56$^\dag$ & 9.09 & 27.84$^\dag$ \\
        4 & & & & 200ms & & 0.18 & 63.81$^\dag$ & 29.15$^\dag$ & 20.19 & 9.31 & 28.08$^\dag$ \\
        5 & & & & 100ms & & 0.10 & 64.49$^\dag$ & 28.36$^\dag$ & 19.47$^\dag$ & 9.27 & 27.87$^\dag$ \\
        6 & & & & 50ms & & 0.06 & 63.99$^\dag$ & 28.68$^\dag$ & 19.17$^\dag$ & 8.99 & 27.70$^\dag$ \\
        7 & & & & 10ms & & 0.03 & \textbf{64.77$^\dag$} & \textbf{29.03$^\dag$} & \textbf{19.21$^\dag$} & 9.09 & \textbf{28.00$^\dag$} \\
    \hline\hline 
        \multirow{3}{*}{\tabincell{c}{DBK.\\Sys.}} & 
        \multirow{3}{*}{\tabincell{c}{Model\\(\#Para.)}} & 
        \multirow{3}{*}{\#Hrs} &
        \multirow{3}{*}{\tabincell{c}{Adapt.\\Feat.}} & 
        \multirow{3}{*}{\tabincell{c}{Slid.\\Wind.}} &
        \multirow{3}{*}{\tabincell{c}{On\\Fly}} &
        \multirow{3}{*}{\tabincell{c}{R.T.\\Factor}} &
        \multicolumn{5}{c}{WER\%} \\
    \cline{8-12}
         & & & & & & & \multicolumn{2}{c|}{Dev} & \multicolumn{2}{c|}{Eval} & \multirow{2}{*}{All} \\
    \cline{8-11}
         & & & & & & & INV & \multicolumn{1}{c|}{PAR} & INV & PAR & \\
    \hline\hline
        1 & \multirow{7}{*}{\tabincell{c}{Hybrid\\TDNN\\(18M)}} & \multirow{7}{*}{58.9} & iVector & 100ms & \multirow{7}{*}{\cmark} & 0.08 & 19.97 & \multicolumn{1}{c|}{46.76} & 18.20 & 37.01 & 33.37 \\
    \cline{1-1} \cline{4-5} \cline{7-12}
        2 & & & \multirow{6}{*}{SVR} & utt. & & 1.03 & 18.72$^\dag$ & \multicolumn{1}{c|}{44.67$^\dag$} & 18.65 & 34.03$^\dag$ & 31.55$^\dag$ \\
        3 & & & & 300ms & & 0.19 & 19.29$^\dag$ & \multicolumn{1}{c|}{45.01$^\dag$} & 19.09 & 33.28$^\dag$ & 31.81$^\dag$ \\
        4 & & & & 200ms & & 0.14 & 19.36 & \multicolumn{1}{c|}{45.15$^\dag$} & 19.20 & 33.89$^\dag$ & 32.01$^\dag$ \\
        5 & & & & 100ms & & 0.08 & 19.08$^\dag$ & \multicolumn{1}{c|}{45.03$^\dag$} & 19.42 & 34.41$^\dag$ & 31.93$^\dag$ \\
        6 & & & & 50ms & & 0.06 & 19.38 & \multicolumn{1}{c|}{44.87$^\dag$} & 18.31 & 34.52$^\dag$ & 31.97$^\dag$ \\
        7 & & & & 10ms & & 0.04 & 19.33$^\dag$ & \multicolumn{1}{c|}{\textbf{44.93$^\dag$}} & 19.09 & \textbf{34.35$^\dag$} & \textbf{31.97$^\dag$} \\
    \hline\hline
    \end{tabular}}
    \vspace{-2mm}
\end{table}

\vspace{-2mm}
\begin{table}[ht]
  \caption{Performance comparison against recently published systems on \textbf{UASpeech}. ``DA'' denotes data augmentation.}
  \label{tab:compare}
  \Large
  \centering
  \vspace{-0.5em}
  \renewcommand\arraystretch{1}
  \scalebox{0.39}{\begin{tabular}{cccc}
  \toprule
    Sys. & Online &VL & All \\
  \midrule
    Sheffield-2015 Speaker adaptive training~\cite{sehgal2015model} & \xmark & 70.78 & 34.85 \\
    Sheffield-2020 Fine-tuning CNN-TDNN speaker adaptation (15spk)~\cite{xiong2020source} & \cmark & 68.24 & 30.76 \\
    CUHK-2020 DNN + DA + LHUC SAT~\cite{geng2020investigation} & \xmark & 62.44 & 26.37 \\ 
    CUHK-2021 LAS + CTC + Meta-learning + SAT~\cite{wang2021improved} & \xmark & 68.70 & 35.00 \\ 
    CUHK-2021 QuartzNet + CTC + Meta-learning + SAT~\cite{wang2021improved} & \xmark & 69.30 & 30.50 \\
    Sheffield-2022 DA + Source Filter Features + iVector adapt~\cite{yue2022acoustic} & \cmark & - & 30.30 \\
    \textbf{DA + SVR Adapt + f-LHUC system combination (Table~\ref{tab:recog-UASpeech}, Sys.15+20)} & \cmark & \textbf{60.80} & \textbf{26.36} \\
  \bottomrule
  \end{tabular}}
  \vspace{-2mm}
\end{table}

\vspace{-0.5em}
\section{Conclusions}
\label{sec-conclusion}
This paper proposes two novel forms of feature-based on-the-fly speaker adaptation approaches: speaker-level variance-regularized spectral basis embedding (SVR) features adaptation and spectral feature driven f-LHUC adaptation. Experiments conducted on benchmark UASpeech dysarthric and DementiaBank Pitt Elderly datasets suggest both methods can efficiently encode homogeneous dysarthric and elderly speaker specific characteristics and outperform both online iVector and offline model based LHUC adaptation. Future research will focus on rapid speaker adaptation of pre-trained ASR systems.

\vspace{-0.5em}
\section{Acknowledgements}
This research is supported by Hong Kong RGC GRF grant No. 14200220, 14200021, TRS T45-407/19N, Innovation Technology Fund grant No. ITS/218/21, National Natural Science Foundation of China (NSFC) Grant 62106255, and  Youth Innovation Promotion Association CAS Grant 2023119.

\bibliographystyle{IEEEtran}
\bibliography{main}

\end{document}